\documentclass{aa}  
\usepackage{graphicx}
\usepackage{txfonts}
\usepackage{float}
\begin{document} 

   \title{Spatially resolved spectroscopy across stellar surfaces. V. }

   \subtitle{Observational prospects: Toward Earth-like exoplanet detection}
  
   \author{Dainis Dravins
          \inst{1},
       Hans-G\"{u}nter Ludwig
           \inst{2},
 \and
       Bernd Freytag
          \inst{3} 
}
%
% Please retain full first names of authors! 
%
\institute{Lund Observatory, Department of Astronomy and Theoretical Physics, Lund University, Box 43, SE-22100 Lund, Sweden\\
              \email{dainis@astro.lu.se}
\and
     Zentrum f\"{u}r Astronomie der Universit\"{a}t Heidelberg, Landessternwarte, K\"{o}nigstuhl, DE--69117 Heidelberg, Germany\\
              \email{hludwig@lsw.uni-heidelberg.de}
\and
     Theoretical Astrophysics, Department of Physics and Astronomy, Uppsala University, Box 516, SE-75120 Uppsala,  Sweden \\
            \email{bernd.freytag@physics.uu.se}
             }
 
\date{Received 26 November, 2020; accepted 1 February, 2021}

% \abstract{}{}{}{}{} 
% 5 {} token are mandatory
 
\abstract
  % context heading (optional)
  % {} leave it empty if necessary  
   {High-precision stellar analyses require hydrodynamic 3D modeling.  Testing such models is feasible by retrieving spectral line shapes across stellar disks, using differential spectroscopy during exoplanet transits.  Observations were presented in Papers~I, II, and III, while Paper~IV explored synthetic data at hyper-high spectral resolution for different classes of stars, identifying characteristic patterns for \ion{Fe}{i} and \ion{Fe}{ii} lines. }
  % aims heading (mandatory)
   {Anticipating future observations, the observability of patterns among photospheric lines of different strength, excitation potential and ionization level are examined from synthetic spectra, as observed at ordinary spectral resolutions and at different levels of noise.  Time variability in 3D atmospheres induces changes in spectral-line parameters, some of which are correlated.  An adequate calibration could identify proxies for the jitter in apparent radial velocity to enable adjustments to actual stellar radial motion. }
  % methods heading (mandatory)
   {We used spectral-line patterns identified in synthetic spectra at hyper-high resolution in Paper~IV from 3D models spanning T$_{\textrm{eff}}$ = 3964--6726\,K (spectral types $\sim$K8\,V--F3\,V) to simulate practically observable signals at different stellar disk positions at various lower spectral resolutions, down to $\lambda$/$\Delta\lambda$ = 75,000.  We also examined the center-to-limb temporal variability. }
  % results heading (mandatory)
   {Recovery of spatially resolved line profiles with fitted widths and depths is shown for various noise levels, with gradual degradation at successively lower spectral resolutions.  Signals during exoplanet transit are simulated.  In addition to Rossiter-McLaughlin type signatures in apparent radial velocity, analogous effects are shown for line depths and widths.  In a solar model, temporal variability in line profiles and apparent radial velocity shows correlations between jittering in apparent radial velocity and fluctuations in line depth. }
  % conclusions heading (optional), leave it empty if necessary 
   {Spatially resolved spectroscopy using exoplanet transits is feasible for main-sequence stars.  Overall line parameters of width, depth and wavelength position can be retrieved already with moderate efforts, but a very good signal-to-noise ratio is required to reveal the more subtle signatures between subgroups of spectral lines, where finer details of atmospheric structure are encoded.  Fluctuations in line depth correlate with those in wavelength, and because both can be measured from the ground, searches for low-mass exoplanets should explore these to adjust apparent radial velocities to actual stellar motion. }


\titlerunning{Spatially resolved spectroscopy across stellar surfaces. V.}
\authorrunning{D. Dravins, H.-G.Ludwig, \& B.Freytag}
\maketitle

\section{Introduction}

This project concerns the atmospheric structure and detailed spectra of solar-type stars, and detection schemes in searches for Earth-like exoplanets in orbits around them.  It aims at recording high-resolution spectra across spatially highly resolved stellar surfaces and to identify signatures from surface features such as granulation or magnetically active regions.  In Papers~I, II, and III of this series \citep{dravinsetal17a, dravinsetal17b, dravinsetal18}, a method for such spatially resolved spectroscopy was elaborated and applied to observations of the G0~V star HD\,209458, and the K1~V star HD\,189733A (``Alopex'').  Differential spectroscopy during exoplanet transits retrieves spectra from the stellar surface portions that temporarily become hidden during successive transit epochs.  Tests of 3D hydrodynamic models of photospheric convection become possible through comparisons to synthetic spectral-line profiles at various center-to-limb positions, computed as temporal and spatial averages over their simulation sequences.  In Paper~IV \citep{dravinsetal21}, photospheric spectral-line shapes and shifts were examined in the 400-700 nm region from complete stellar spectra synthesized at hyper-high spectral resolution ($\lambda$/$\Delta\lambda>$1,000,000), obtained from 3D hydrodynamic CO\,$^5$BOLD simulations \citep{freytagetal12, ludwigetal21}.  These spectral signatures encode the detailed properties of the dynamic stellar photosphere, and can serve as a diagnostic tool for testing and verifying hydrodynamic simulations for various classes of stars.  Sufficiently developed models should be able to predict stellar microvariability in radial-velocity jittering and to identify possible proxies, such as could be measured in parallel in order to segregate stellar surface variations from its center-of-mass wobble induced by small orbiting exoplanets.

This Paper~V examines the practical observability of specific spectral features (and their variability in time) when observed with realistic instrumentation.  It is organized as follows: After the first introductory parts, Sect.\ 3 illustrates how successively lower spectral resolutions affect the observable spectral line shapes and shifts, Sect.\ 4 simulates the retrieval of spatially resolved spectra from exoplanet transit sequences at various noise levels, Sect.\ 5 shows the signatures of the Rossiter-McLaughlin type that exist in the remnant stellar light during an exoplanet transit, Sect.\ 6 examines the short-term temporal variability to identify spectral-line parameters that correlate with jittering in apparent radial velocity, and Sect.\ 7 outlines challenges and problems that have not yet been treated. 

\section{Spectral lines from realistic 3D models}

While the overall feasibility of spatially resolved spectroscopy and the detection of specific 3D signatures was demonstrated in Papers~I, II, and III, the finer details of hydrodynamic atmospheric structure are encoded in the different responses in various types of lines, where the practical observability is less obvious.  Paper~IV exploited noise-free synthetic spectra at hyper-high resolution to understand how many of these characteristic patterns in \ion{Fe}{i} and \ion{Fe}{ii} lines remain detectable in also realistically complex stellar spectra, where practically all blending and other lines are included.  Line profiles and bisector shifts between stronger and weaker lines reflect properties at different atmospheric heights and change between stars with different gradients in their line-forming volumes.  Higher-excitation lines may preferentially form in hotter regions than low-excitation lines although components of neutral atomic species from the hottest locations may be weakened due to deviations from local thermodynamic equilibrium (NLTE effects).  There, increased ultraviolet flux can ionize the line-forming layers above; in some cases, there are prospects for studying these effects separately for the hotter granules and the intergranular lanes.  Time variability from 3D atmospheres implies correlations between various spectral-line measures, suggesting further detailed tests of 3D models.  

\subsection{The quest for Earth-like exoplanets}

An outstanding challenge is to find closely Earth-like exoplanets, whose sizes, masses, and orbits are comparable to the terrestrial values.  One plausible path toward their detection is to identify the wobble in radial velocity caused by the stellar motion around the system barycenter, induced by the orbital motion of the planet(s).  Although the tiny amplitude induced by the Earth on the Sun amounts at most to only 0.1 m\,s$^{-1}$ \citep[e.g.,][]{halletal18}, instrument precisions are now beginning to reach these levels \citep{blackmanetal20, breweretal20, caleetal19, gilbertetal18, langellieretal20, locurtoetal12, metcalfetal19, pepeetal14, petersburgetal20, probstetal20, robertsonetal19, royetal16, strassmeieretal18a, suarezmascarenoetal20, wilkenetal12, zhaoetal21} and they are design criteria for future instruments \citep{halversonetal16, pasquinietal10, plavchanetal19, zerbietal14}.  The main limitations are no longer technical but lie in understanding the complexities of atmospheric dynamics and spectral line formation, manifest both as a jitter of the apparent radial velocity and as a flicker in photometric brightness \citep{cegla19, fischeretal16}. 

In the Sun and solar-type stars, with maybe a million granules covering their disks, wavelength positions of photospheric absorption lines jitter on a level of perhaps $\sim$2 m\,s$^{-1}$ \citep{colliercameronetal19, dumusqueetal15, dumusqueetal20}, an order of magnitude that is consistent with the averaged random variability of a million granules, each with an amplitude of $\sim$2 km\,s$^{-1}$.  Photometric flickering in irradiance is analogous: A typical brightness contrast of $\sim$20\,\% in each granule reduces to $\sim$2\,10$^{-4}$ after averaging over a million random locations \citep[e.g.][]{nemecetal20}.  The amplitude diminishes with increased averaging times \citep{bastienetal16, cranmeretal14}.  Further phenomena such as starspots or faculae are adding to the confusion \citep{isiketal20, korhonenetal15, lagrangeetal10, lisogorskyietal20, meunieretal10, milbourneetal19, norrisetal17}.
 
The radial-velocity wobble induced by a small exoplanet is much smaller than stellar microvariability, therefore effects of the latter must somehow be calibrated and corrected for.  Numerous authors have studied empirical correlations between various stellar activity indices and corresponding excursions in apparent radial velocity \citep[e.g.,][]{ceglaetal14, ceglaetal19, colliercameronetal20, debeursetal20, dumusque16, fengetal16, haywoodetal16, hojjatpanahetal20, kosiarekcrossfield20, lanzaetal19, luhnetal20a, marchwinskietal15, meunieretal10, meunieretal15, meunieretal17b, miklosetal20, oshaghetal17, sulisetal20a, tayaretal19, thompsonetal20}.

Most studies have considered the Sun seen as a star, where activity patterns on the visible disk are known from observations from the ground (e.g., sunspots, magnetic fields, and chromospheric Ca II H \& K emission) or from space (photometric irradiance and ultraviolet flux).  Numerous techniques have been applied, analyzing mathematical correlations between various quantities or applying neural networks.  Temporal filtering can distinguish between evolutionary and recurrence timescales, for instance, active-region growth times versus stellar rotation periods.   However, even if solar variations can thus be traced and a mitigation of radial-velocity signals is reached, the same means are not available for other stars.  Relations between stellar photometric flicker and wavelength jitter confirm the impact of magnetic activity on apparent radial velocities but the relations vary among different stellar types and depend on metallicity \citep{tayaretal19}.

Although various solar-calibrated relations may be applied to also other stars, these correlations do not yet reach sufficient predictability to permit the identification of low-mass exoplanets in Earth-like orbits.  We suspect that such types of empirical correlations may not be able to approach the required precisions, and a different approach would therefore be needed. 

This could be based upon a detailed physical modeling of temporal variability in stellar atmospheres, precisely predicting how various types of thermal and magnetic variability are manifest in the details of the emergent and observable spectrum.  It is enabled by the recent feasibility to simulate entire and complete time-variable stellar spectra including practically all spectral lines, and possible for different classes of stars.  

If these processes can be adequately understood from more basic principles, it should become possible to identify which spectral parameters in what spectral lines should best be observed in different stars, and what observational precision will be required to identify small exoplanets as well.  Recent efforts in this direction for the Sun seen as a star include the extraction of radial-velocity data from separate sublists of lines with different depths \citep{miklosetal20} and the study of how extended spectral portions differ at epochs of magnetically more or less active periods \citep{thompsonetal20}. 

Fluctuations in hydrodynamic atmospheres affect many spectral-line parameters in concert.  Line components originating from hotter granules in their normally rising motion become locally blueshifted and carry higher photon fluxes due to brighter continuum intensities.  Steeper gradients in density and temperature above granules tend to cause increased strengths of absorption lines.  Lower opacity in cooler intergranular lanes may enable lines to form over greater depths, and to experience vertical velocity gradients, which leads to more pronounced line asymmetries.  All these dependences may further vary among lines with different temperature sensitivities, for example, high-excitation lines from ionized species versus lines from easily dissociated molecules \citep[e.g.,][]{ allendeetal02, chiavassaetal18, dravinsnordlund90, nordlundetal09, trampedachetal13}. 

If it can be adequately understood which parameters can serve as proxies for jittering in radial velocity (and especially if these quantities can be measured from the ground), it should enable the push toward sub-m\,s$^{-1}$ precisions, as required for exoEarth detections.  It is clear that there is additional information content in individual spectral lines, as compared to statistically averaged correlation functions and such \citep{dumusque18, cretignieretal20}, but how much that can be retrieved must depend on how precisely each particular spectral line or line group can be measured.  Various correlations among parameters may be searched from extended 3D simulations and could be verified in low-noise observations of bright stars. When these correlations are understood, they might be used in radial-velocity searches for low-mass exoplanets to adjust measured radial-velocity values to true stellar radial motion, at least for the displacements caused by thermally driven granulation.  Additional complexities are to be considered separately, in spatially resolved spectroscopy of magnetically active regions and of starspots.

\subsection{Transit light source effects} 

A further motivation for studying spatially resolved stellar spectra is that atmospheric properties of any transiting exoplanet must be deduced as subtle differences against the background stellar spectrum, which thus must be precisely known.  This requires a knowledge of the varying stellar line profiles at the positions along the transit path of the planet \citep{apaietal18, rackhametal18, rackhametal19, yanetal17}; these are the features that are filtered through the exoplanetary atmosphere, not the spectrum of the flux from the full stellar disk.  Additionally, the spectrum of the stellar flux that remains during exoplanetary transit (then displaying the Rossiter-McLaughlin effect) depends not only on stellar rotation but is also modified by the change in line profiles across the stellar disk.

\begin{figure}
 \centering
 \includegraphics[width=\hsize]{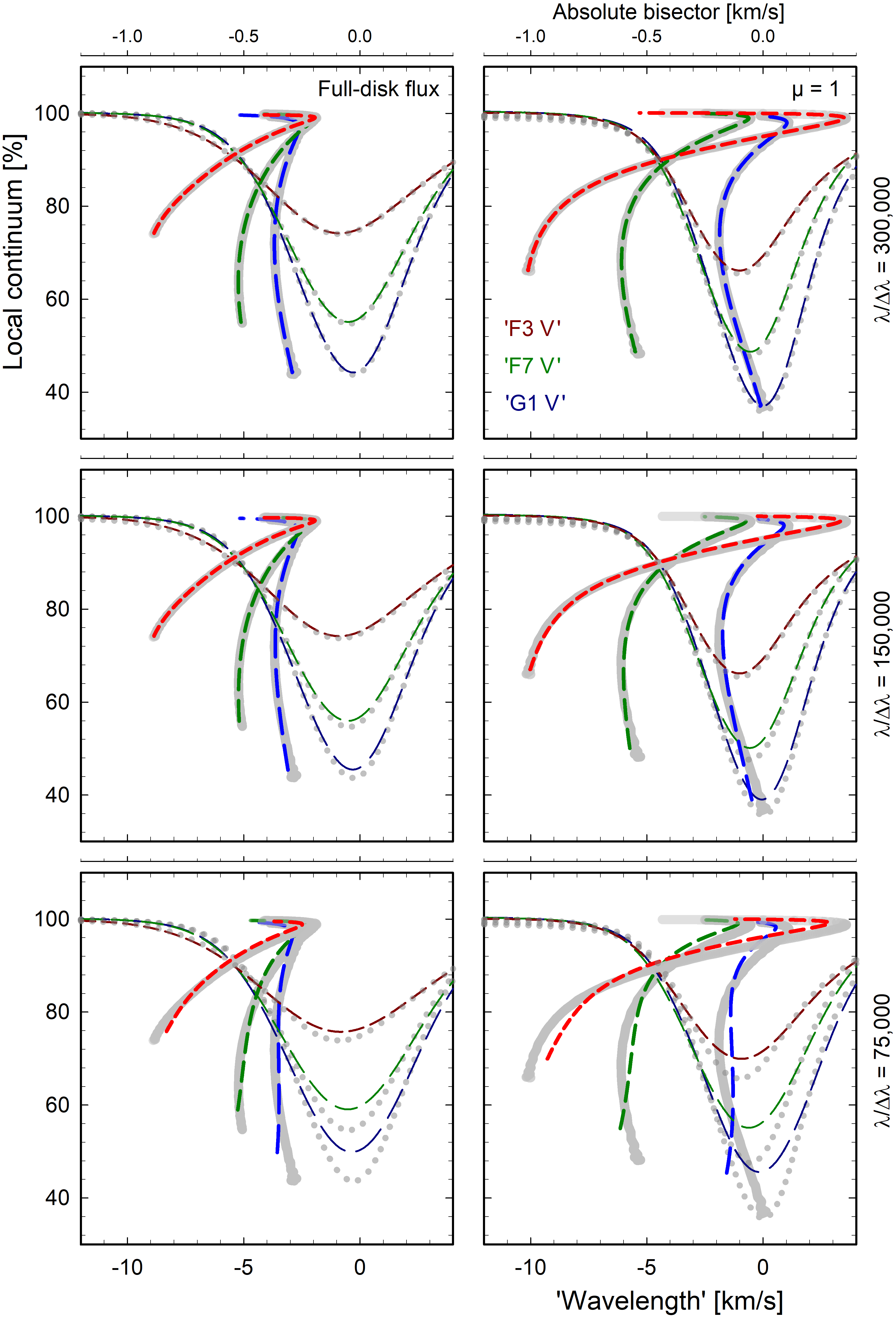}
     \caption{Effects of decreasing spectral resolution on line profiles (axes at the bottom) and their bisectors (tenfold expanded axes at the top) for a representative spectral line.  From the top down, three rows show the \ion{Fe}{i} $\lambda$ 539.8286 nm line for $\lambda/\Delta \lambda$ = 300,000, 150,000, and 75,000.  The left column shows data for integrated full-disk flux, and the right column that for stellar disk center ($\mu$ = 1).  The same \ion{Fe}{i} line is shown in models for `F3~V', `F7~V', and `G1~V' as short-, medium-, and long-dashed lines (red, green, and blue).  Each frame also shows profiles and bisectors at the original hyper-high spectral resolution, marked as dotted gray and solid gray curves. } 
\label{fig:different_resolutions}
\end{figure}

\section{Finite spectral resolution}
 
All data in Paper~IV were at the original hyper-high spectral resolutions $\lambda$/$\Delta\lambda$ $>$1,000,000.  Although these levels are required to (almost) fully resolve line asymmetries and shifts, this will likely not be available for stellar observations in the near future.  In this section we examine the spectral-line signatures that are preserved at the lower spectral resolutions that are commonly achievable.

Spectra at an order of magnitude lower resolution of $\lambda$/$\Delta\lambda$ $\sim$100,000 may be adequate to identify the existence of almost all spectral lines but not to reveal their precise line shapes.  The effects of finite resolution can be appreciated by noting that because the relevant width of a typical photospheric line may be about 12~km~s$^{-1}$,  a resolution of 100,000 (3~km~s$^{-1}$) provides four resolution points across that line profile, while a resolution of 300,000 (1~km~s$^{-1}$) gives 12 points.  A bisector is obtained by averaging pairs of resolution points in the shortward and longward flanks of the line.  The number of data points on the bisector is thus half of these numbers, in these examples 2 and 6, respectively.  With very few independent points, we cannot define much more than a straight line, indicating the bisector wavelength position and its slope (i.e., the sense of line asymmetry).  To detect any curvature on a bisector requires several photometric measurements across the relevant portions of a line.  The exact requirements depend on the intrinsic width of the lines, and are less stringent for broader lines (such as the line in Fig.\ \ref{fig:different_resolutions}).  The degradation of bisector shapes with dropping resolving power is well illustrated for observed solar lines when the resolution is gradually decreased from 700,000 to 100,000 \citep{lohnerbottcheretal18, lohnerbottcheretal19}. 

Any spectral synthesis is limited not only by computational sophistication but also by the (non-)availability of precise laboratory data for spectral lines from all species, including components from different isotopes or hyperfine structure patterns.  A circumvention of these problems in comparing modeling with observations suggests preferentially studying atomic species that are dominated by one isotope that is even-even in its proton-neutron numbers, so that its nuclear spin and thus hyperfine splitting is zero.  In addition, this species should be of high mass, minimizing the thermal broadening of the lines, to better segregate effects of photospheric gas flows.  Similar to other studies, the preferred atomic species is iron.  Its dominant isotope, $_{26}$Fe$^{56}$ has a (terrestrial) abundance of 91.7\% while a few isotopes with lighter and heavier mass exist: $_{26}$Fe$^{54}$ has 5.8\%, $_{26}$Fe$^{57}$ has 2.1\%, and $_{26}$Fe$^{58}$ has 0.3\% \citep{delaeteretal03}.  The nuclear spin $I$ and hyperfine splitting are zero for $_{26}$Fe$^{54}$, $_{26}$Fe$^{56}$, and $_{26}$Fe$^{58}$ and small for $_{26}$Fe$^{57}$ ($I = \frac{1}{2}$).  Thus, asymmetries induced by isotope shifts and hyperfine structure should be negligible.  When adequate laboratory data and ensuing modeling become available, it should be possible to extract additional information from the detailed behavior of lines from species whose hyperfine splitting spans ranges corresponding to photospheric velocity amplitudes; for instance, patterns for $_{25}$Mn$^{55}$ of 100\% abundance and $I = \frac{5}{2}$, extend over $\sim$10 pm ($\sim$5~km~s$^{-1}$).  In the following, however, we limit the discussion to the \ion{Fe}{i} and \ion{Fe}{ii} line selections described in Paper~IV.

\begin{figure*}
%\sidecaption
\centering
 \includegraphics[width=16.1cm]{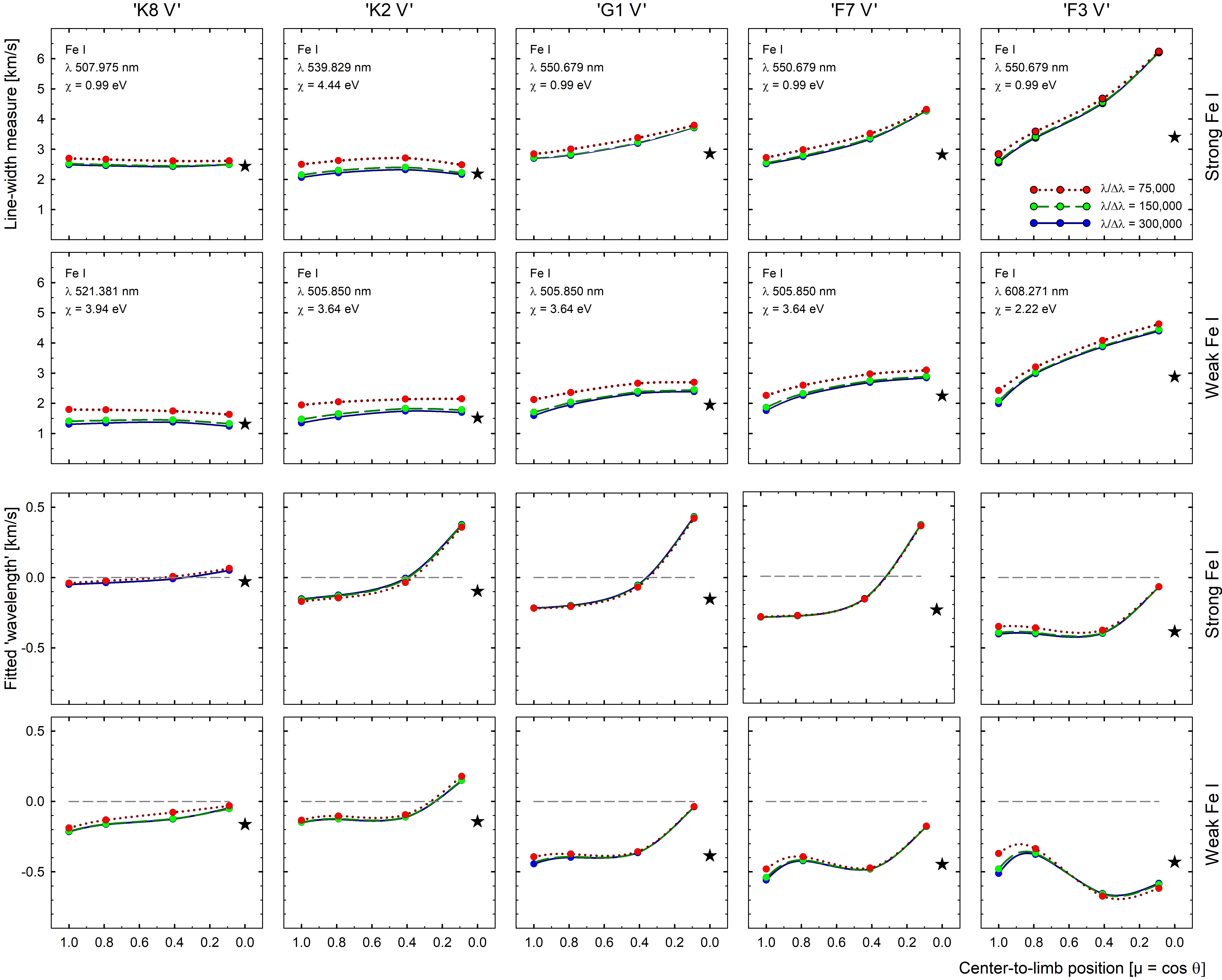}
     \caption{Parameters obtained from fitting full line profiles at various spectral resolutions in different stars.  Top rows: Fitted line widths for one weak and one strong line at different stellar disk positions.  Bottom rows: Corresponding wavelength shifts.  The values were obtained from fitting five-parameter Gaussian-type functions to the line profiles.  Black stars denote values for integrated full-disk flux from 3D models at full hyper-high spectral resolution.  Solid curves with blue symbols denote $\lambda/\Delta \lambda$ = 300,000, dashed lines with green symbols show 150,000, and dotted curves with red symbols plot the lowest resolution of 75,000. } 
     \label{fig:resolution_parameters}
\end{figure*}

\subsection{Observational possibilities}

Anticipating future observations, we now examine the signatures in spectral lines that might realistically be observed at resolutions corresponding to those of actual stellar spectrometers at major telescopes.  Several precision instruments realize $\lambda$/$\Delta\lambda$ $\sim$100,000; resolutions of about 200,000 are possible with ESPRESSO at the ESO VLT\footnote{ESPRESSO, the Echelle SPectrograph for Rocky Exoplanets and Stable Spectroscopic Observations, at VLT, the Very Large Telescope of ESO, the European Southern Observatory} \citep{pepeetal14, pepeetal21, gonzalezhernandezetal18} but extended spectral ranges approaching $\lambda$/$\Delta\lambda$ $\sim$300,000 are currently only offered by PEPSI at the LBT\footnote{PEPSI, the Potsdam Echelle Polarimetric and Spectroscopic Instrument at LBT, the Large Binocular Telescope} \citep{strassmeieretal15, strassmeieretal18b}.  Actual performance comparisons have been made by \citet{adibekyanetal20}, \citet{benatti18}, and \citet{crossfield14}.  Despite the ample photon fluxes in future extremely large telescopes, conventional spectrometer designs will likely not reach equally high values, constrained by the difficulty to match diffraction grating sizes to the large image scales.  This holds for HIRES on the ELT\footnote{HIRES, the High Resolution Spectrograph planned for ELT, the Extremely Large Telescope of ESO} \citep{marconietal16, marconietal21, zerbietal14} or concepts such as CODEX\footnote{CODEX, COsmic Dynamics and EXo-earth experiment, a spectrograph originally proposed for the then OWL, Overwhelmingly Large Telescope, project} \citep{pasquinietal10}.  Spectrometers with adaptive optics fed by single-mode fibers might be able to circumvent this limitation \citep{bechteretal18, beckersetal07, crassetal19, jovanovicetal19}.  A few other instruments have occasionally been used to record limited spectral segments at even higher resolutions, but they are not adapted to extended spectral regions.  The resolution values typically refer to the instrumental profile full width at half maximum, which the detector normally samples with at least two detector pixels.  

Our models have the temperatures T$_{\textrm{eff}}$ = 3964, 4982, 5700, 5865, 6233, and 6726\,K, and similar to Paper~IV (its Table~1), these are referred to with their approximate corresponding spectral types as `K8~V', `K2~V', `G2~V', `G1~V', `F7~V' and `F3~V'.  Fig.\ \ref{fig:different_resolutions} shows the successive degradation of one strong line (\ion{Fe}{i} $\lambda$ 539.8286 nm) when shifting from hyper-high resolution to the observationally more realistic values of 300,000, 150,000 and 75,000.  Three stellar models are shown for integrated full-disk spectra and for the stellar disk center.  Most of the line signatures are preserved at $\lambda/\Delta \lambda$ = 300,000 but below $\sim$150,000, line asymmetries become distorted (as expected). 

Observed full-disk spectra will display additional broadening arising from (rigid or differential) stellar rotation and possible large-scale surface oscillations or flows.  Spatially resolved spectra are free from these effects and will instead be shifted in wavelength by amounts corresponding to the components of stellar rotation along the line of sight at each particular position of the transiting planet (Papers~II and III).  We recall that the hydrodynamic simulations do not entail any fitting parameters.  In particular, concepts such as turbulence do not enter -- spectral line shapes result from their formation within the computed volumes of gas flows at different optical depths along the line of sight.

\subsection{Fitting line parameters}

In realistic observations, noise levels constrain the detailed analysis of line profiles and only some line-averaged measures might be obtained from some fitting, smoothing, or averaging.  Fig.\ \ref{fig:resolution_parameters} shows how such fitted widths and shifts are altered, depending on the achieved spectral resolution.  As an approximation, the profiles (in this section still assumed to be noise-free) at each spectral resolution were fit with five-parameter Gaussian-type functions of the type $y_0 + a\cdot\exp[-0.5\cdot(|x-x_0|/W)^c]$, yielding $W$ as a measure of the line width, and $x_0$ for the central wavelength.  Although the profiles are not precisely Gaussian, these simple functions appear reasonable to use in fitting imperfect data.  Different functions have previously been tested (Lorentzian, etc.), but it was found that plain Gaussian varieties gave robust estimates for the basic line parameters.  A fitting like this is relevant in cases of noisy data when full line profiles cannot be retrieved, and then only simpler fitting functions are likely to be justified.  The parameters of line-width measure and (average) wavelength position appear to be the most easily measurable quantities, also for the center-to-limb dependences in spatially resolved spectra.  In particular, for many stars, the line widths are expected to increase toward the limb, providing a quantitative measure that the greater horizontal motion amplitudes contribute more Doppler broadening.  

In Fig.\ \ref{fig:resolution_parameters} it is striking that much of the signatures also remains at lower spectral resolutions.  When the value has reached $\lambda$/$\Delta\lambda$ $\sim$150,000, extending to even higher resolutions brings only very slight changes in parameters that are fit to the overall line profiles.

The data in Fig.\ \ref{fig:resolution_parameters} are for representative lines of different strength, which is the one factor (rather than excitation potential or wavelength region) that appears to be foremost in affecting the spectral parameters.  Differently strong lines are prevalent in different parts of the spectrum: statistics of atomic energy levels dictate that most strong lines from neutral metals appear in the blue or violet parts, while weaker transitions are common in the red and infrared.  In general, convective signatures can be expected to be weaker at longer wavelengths because any temperature difference between thermally radiating hot and cool elements causes a lower intensity contrast in the (infra)red, diminishing the statistical imbalance between blueshifted line components from hot granules and redshifted contributions from intergranular regions \citep[e.g.,][]{asplundetal00}.  A caveat is that specific signatures might appear due to locally different atmospheric opacity in specific regions, however.

Single numbers for line width and wavelength are not uniquely defined measures, however, but depend on exactly how the fit is made, and with what type of weighting.  For example, a wavelength measure may depend on whether the value is obtained from a fit to the central or more extended parts of the profile, or perhaps made with some correlation template that gives increased weight to steeper intensity gradients.  For the more general question of quantifying spectral resolution in astronomical spectra, see \citet{robertson13}.

\begin{figure*}
\sidecaption
 \centering
 \includegraphics[width=12.9cm]{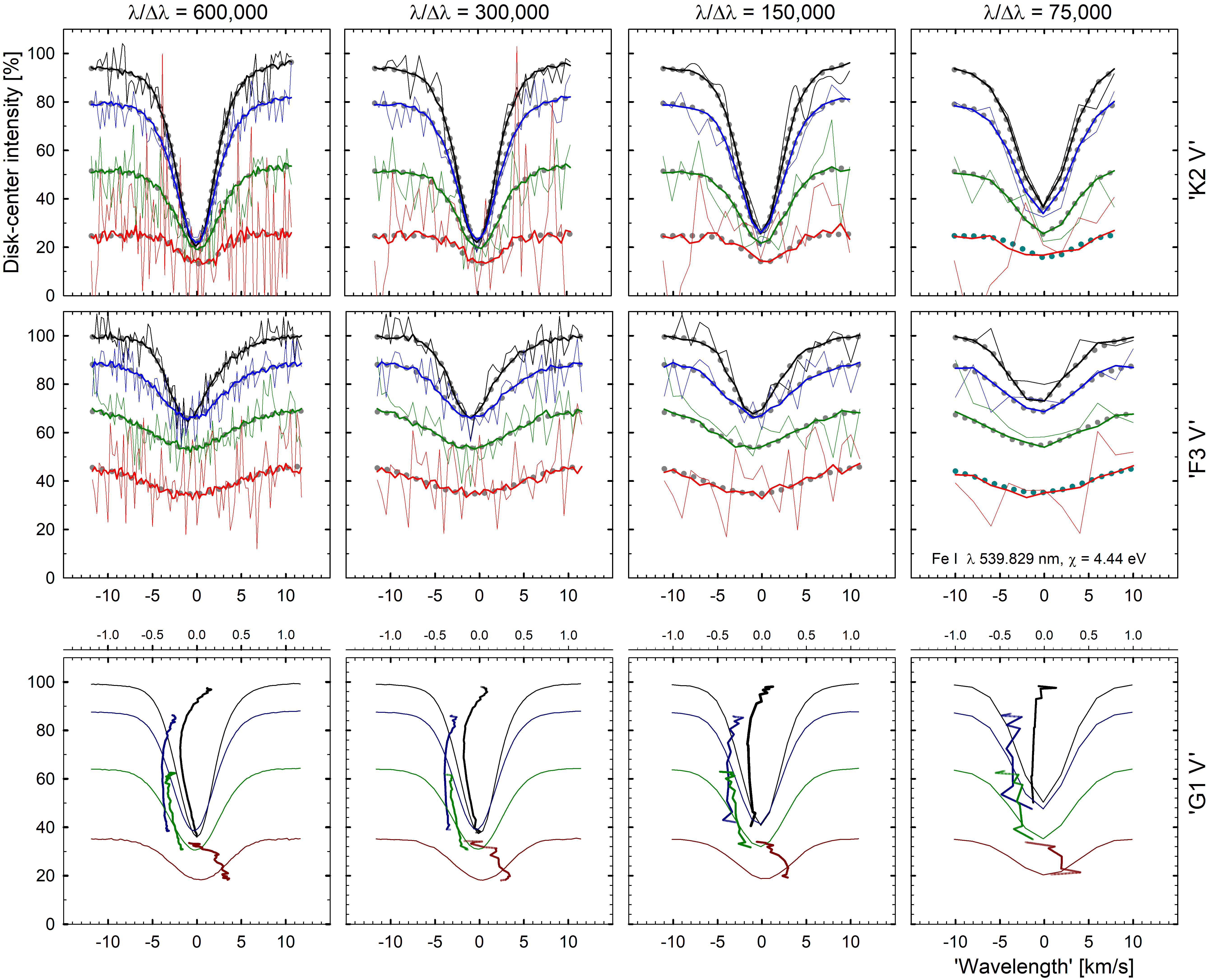}
     \caption{Simulated noise in spatially resolved line profile reconstructions from observations during an exoplanet transit.  Columns from left to right show reconstructions from resolutions $\lambda$/$\Delta\lambda$ = 600,000, 300,000, 150,000 and 75,000, sampled with two pixels per spectral resolution element.  The two top rows show a cooler `K2~V' and a hotter `F3~V' star, differing in line strengths, widths, and limb darkening.  Photometric noise levels during transit in the spectral continuum are: $\sigma$ = 0.01\% (thin line) and 0.001\% (bold).  Noise-free profiles at each resolution are plotted as gray dots.  Inside each frame and top down, profiles are for disk center positions $\mu$ = 1, 0.79, 0.41, and 0.09.  Bottom row: Reconstructed profiles at the extremely low noise level of $\sigma$ = 0.0001\%, which is required to also recover well the bisectors (tenfold expanded scale at top). } 
     \label{fig:noisy_reconstruction}
\end{figure*}

\section{Reconstructed profiles from noisy observations}

To reconstruct high-resolution spectra from behind the small areas covered by a transiting planet requires signal-to-noise ratios (S/N) much higher than commonly available in individual spectral exposures.  As detailed in Papers I, II, and III, the requirements limit usable observations to spectra with the highest-fidelity from high-resolution spectrometers.   For example, a S/N in the reconstructed spectrum of about 100, extracted from $\sim$1\% of the total stellar signal, requires an original S/N of about 10,000.  The remedy applied in Papers~II and III was to form averages over numerous lines of similar strength that form under similar conditions in the stellar atmosphere, and are expected to display similar characteristics.  These possibilities, however, may be limited to certain spectral regions or to cooler stars with particularly rich spectra.  

To appreciate the levels of spectral quality that can be realized at different levels of observational noise, simulated observations were calculated at different spectral resolutions and for noise levels bracketing values that provide meaningful reconstructions.  The area of a transiting planet, $A$, was assumed equal to 1.5\% of the stellar disk for each stellar model, roughly corresponding to a Jupiter-size object.  At each disk position, the temporarily hidden spectral signal $S$ equals that from the full stellar disk (outside transit) minus that from the unobscured portions, its normalization in intensity set by $A$; thus: $S_{\textrm{hidden}}$ = ($S_{\textrm{outside-transit}} - S_{\textrm{in-transit}})/A$.  Because the reconstructions involve dividing subtracted spectra by the low values of $A$, the noise is correspondingly enhanced.  For a more detailed discussion of the steps in reconstructing spatially resolved profiles, see Paper~II (e.g., its Fig.\,9).  Spectral reconstructions like this were here simulated from observations at spectral resolutions $\lambda$/$\Delta\lambda$ = 600,000, 300,000, 150,000 and 75,000, sampled by spectrometers with two pixels per spectral resolution element.  Random photometric noise was assumed with its standard deviation proportional to the square root of the flux in each pixel, and for exposures during transit, set to $\sigma$ = 0.01\%, 0.001\%, and 0.0001\% of the spectral continuum.  Because the stellar reference spectrum from outside the transit can be much better determined than those from brief exposures during transit, its noise level was set a factor of ten lower.  These simulated line-profile reconstructions are shown in Fig.\ \ref{fig:noisy_reconstruction}

Although the same line, \ion{Fe}{i} $\lambda$ 539.82860 nm, was used for all three stellar models `K2~V', `G1~V', and `F3~V', the greater limb darkening for lower stellar temperatures has a certain effect on the profile reconstructions away from disk center because the signal from a planet away from disk center is weaker.  The noise levels of 0.01\% and 0.001\%, in the top rows of Fig. \ref{fig:noisy_reconstruction} illustrate the range within which meaningful reconstructions of line profiles can be performed.  These are adequate to obtain line-integrated parameters such as widths, depths and wavelength shifts, even if line profile shapes were lost (cf.\ Fig.\ \ref{fig:resolution_parameters}).  The actual reconstructions in Papers~II and III were made at roughly comparable (or slightly worse) noise levels.  The more demanding recovery of detailed line profile shapes and their precise bisectors will require both a very high spectral resolution and an extremely low noise level, as illustrated in the bottom row for a continuum noise $\sigma$ = 0.0001\% (S/N = 10$^6$), which might be approachable after extensive stacking of lines and exposures using future instruments on extremely large telescopes. 

The role of different planet sizes should be recognized.  Realistic line-profile reconstructions demand transit observations of large planets, for instance, Jupiter-size, covering $\sim$1\% of the disk of solar-type stars.  Not only is this to satisfy the spectrophotometric requirements but also to provide adequate averaging over stellar surface inhomogeneities because such a planet covers $\sim$10,000 granules out of the total $\sim$10$^6$ granules that are present on such stars.  A smaller planet would give not only a weaker signal in its differential spectroscopy between transit epochs but also generate more astrophysical noise because the averaging area across stellar surface structures is smaller.  

We recall that a plausible method for finding small planets is to try to identify the radial-velocity wobble of the host star as it moves in a reflex motion with the planet orbiting their common barycenter.  Regardless of instrumental precision, such radial-velocity data have to be corrected to segregate fluctuations arising in the stellar atmosphere from the smaller variations induced by a possible small planet.  Spatially resolved spectra retrieved from transits of large planets can be used to verify and constrain 3D simulations and these models can then be used to predict and calibrate spectral signatures of the host star, such as its apparent radial-velocity variations.  However, it is not realistic to observationally retrieve stellar atmospheric properties from transits of Earth-size planets, covering only $\sim$10$^{-4}$ of the stellar disk. In the following sections, we consider various types of simulated observations during, and outside planetary transits.

\section {Rossiter-McLaughlin-type signatures}

During the transit of an exoplanet across a rotating star, it selectively hides different portions of the stellar surface, where the local rotational velocity vector is toward or away from the observer.  Thus, part of the blue- or redshifted photons are removed from the integrated stellar flux, whose averaged wavelength then appears slightly red- or blueshifted.   The effect has been measured for numerous exoplanets, where measurements of this Rossiter-McLaughlin effect enable determining the geometry of the projected path of the exoplanet \citep{albrecht12, perryman18, winn11}.

\begin{figure*}
% \centering
\sidecaption
 \includegraphics[width=12.9cm]{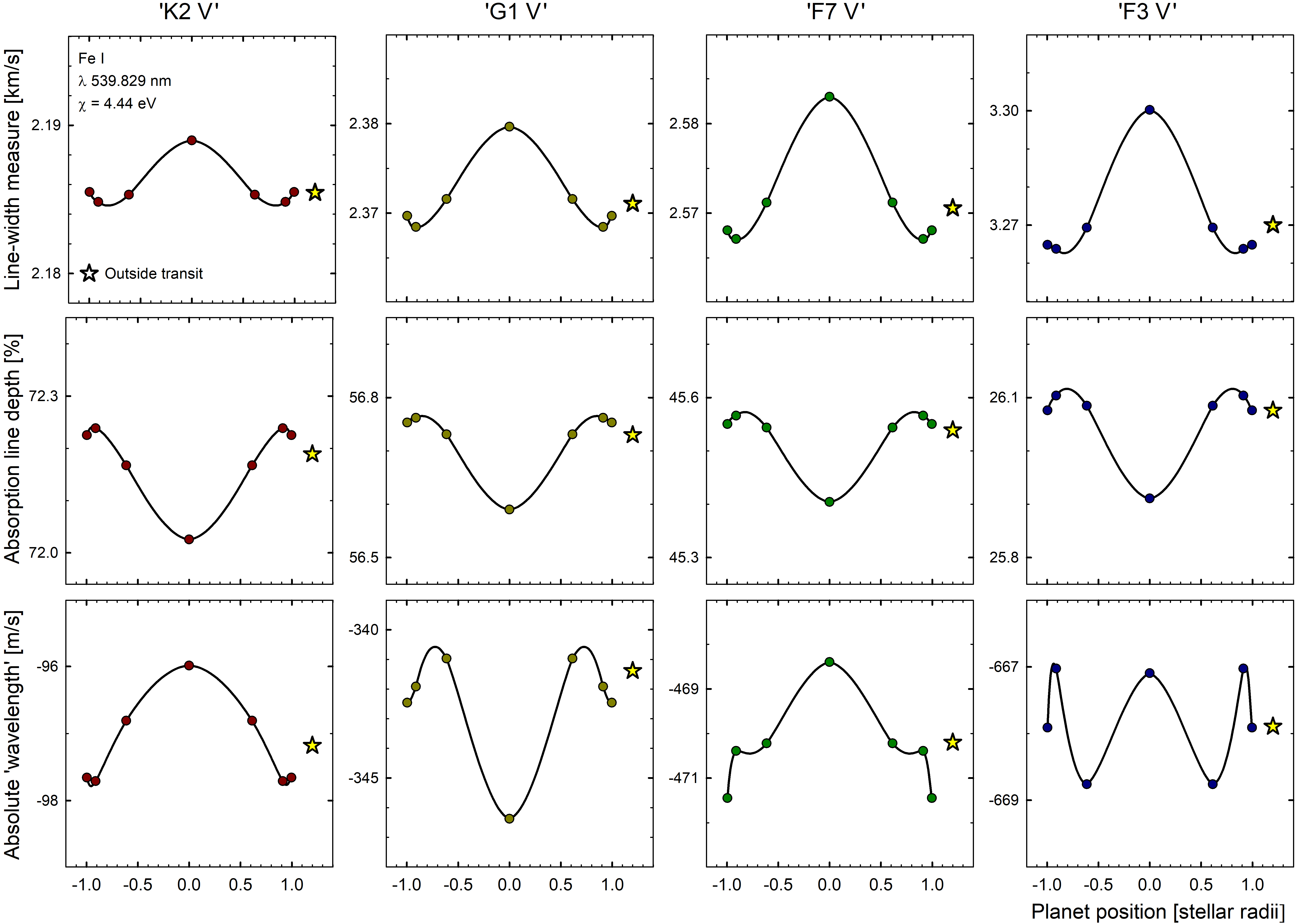}
     \caption{Rossiter-McLaughlin-type signatures during the transit across a stellar diameter of an exoplanet covering 1.5\% of the stellar disk, caused by center-to-limb changes in line profiles in the absence of stellar rotation.  Columns left to right show the `K2~V', `G1~V', `F7~V', and `F3~V' models.  Values for spectral-line widths, depths and wavelength shifts were obtained by fitting \ion{Fe}{i} $\lambda$ 539.8286 nm profiles with the planet at different disk positions.  Line depths are in units of the local spectral continuum at each planetary transit position. Star symbols show full-disk values outside transit.  Wavelength shift values are absolute but gravitational redshifts are neglected. } 
 \label{fig:rm-signatures}
\end{figure*}

However, in addition to the line displacements due to obscured line contributions shifted by stellar rotation, a contribution comes from the intrinsic center-to-limb changes of stellar wavelengths caused by convective shifts, as previously evaluated by \citet{ceglaetal16}, \citet{reinersetal16}, \citet{shporerbrown11} and \citet{yanetal15a, yanetal15b, yanetal17}.  In Fig.\ \ref{fig:rm-signatures} we show detailed calculations for one line (the same as in Fig.\ \ref{fig:different_resolutions}), for four different stellar models, not only for this usual wavelength displacement but also for the concurrent changes in line widths and depths.  A Jupiter-class exoplanet, covering 1.5\% of each stellar disk was assumed, and profiles of the \ion{Fe}{i} $\lambda$ 539.8286 nm were fit with five-parameter Gaussians as above, in the resulting spectrum of the full-disk flux minus what is obscured by the planet at each respective disk position.  While this example is for a rather strong \ion{Fe}{i} line at the full hyper-high spectral resolution, the patterns are quite similar at lower resolutions as well, and also for other lines of comparable strength.  The wavelength shifts are absolute but gravitational redshifts are neglected.  The qualitative behavior can be readily understood: Lines are narrowest and deepest at disk center and when the planet obscures this region, the line width of the remaining stellar flux becomes broader, and the line depth shallower.  The run of wavelength shifts is more complex because line asymmetries and convective line shifts are not necessarily monotonic with limb distance (Fig.\ \ref{fig:resolution_parameters}).

\section{Temporal flickering and jittering}

The signatures discussed so far have been temporal averages but further details about stellar surfaces are encoded in their temporal variability.  Stellar surface dynamics cause flickering in irradiance and photometric colors, and jittering in spectral line wavelengths, widths, shapes and depths.  Although light from full stellar disks forms as the mean over perhaps a million granules, averaging over random structures decreases the signal by only the square root of their number, a slowly changing function.  Full-disk flickering in broadband light retains amplitudes that are measurable by precision photometry from space.  In temporal power spectra, granulation is detected on characteristic timescales, different from those of oscillations or stellar rotation, for instance.  Details depend on stellar spectral type, are possible to model, and inversely, can be used to infer stellar properties from the observed flickering \citep{bastienetal16, cranmeretal14, kallingeretal14, lundkvistetal21, pandeetal18, samadietal13a, samadietal13b}.  Understanding full-disk flickering is also essential to appreciate the noise floor for exoplanet transits in current and future space photometry missions \citep{morrisetal20a, sarkaretal18, sulisetal20b}.

Sharper diagnostics for the finer details of stellar atmospheres exist in the temporal variability of spectral lines, especially versatile when simultaneous fluctuations can be monitored in different types of lines.  Variability that does not require data from space but can be recorded with instrumentation from the ground is particularly relevant.  During instances when there happen to be particularly many bright granules on the stellar disk, more contributions of blueshifted line components from rising gases in the bright granules are expected, causing an enhanced convective blueshift.  At the same time, the augmented irradiance implies an increased temperature and a bluer photometric color.  Locally augmented irradiance may imply locally strengthened atmospheric gradients with ensuing changes to spectral line strengths.  In this situation, photometric brightness, color, or some spectral-line parameter might be used as a proxy for excursions in apparent radial velocity and then be applied in order to adjust it to the physical stellar motion.  To observationally explore and confirm these correlated relations, very bright stars where low-noise spectra can be obtained, would need to be observed first before the relations would be applied to exoplanet hosts.

\subsection{Line profile variability} 

We examined the short-term spectral-line variability as computed from the temporal evolution in the simulation volumes.  Eventually, it should become possible to study the temporal wavering of complete stellar spectra and to identify which line selections, spectral filters, correlation functions or other measures could best extract specific information on processes in diverse atmospheric structures at different atmospheric heights.  

\begin{figure*}
\sidecaption
 \centering
 \includegraphics[width=12cm]{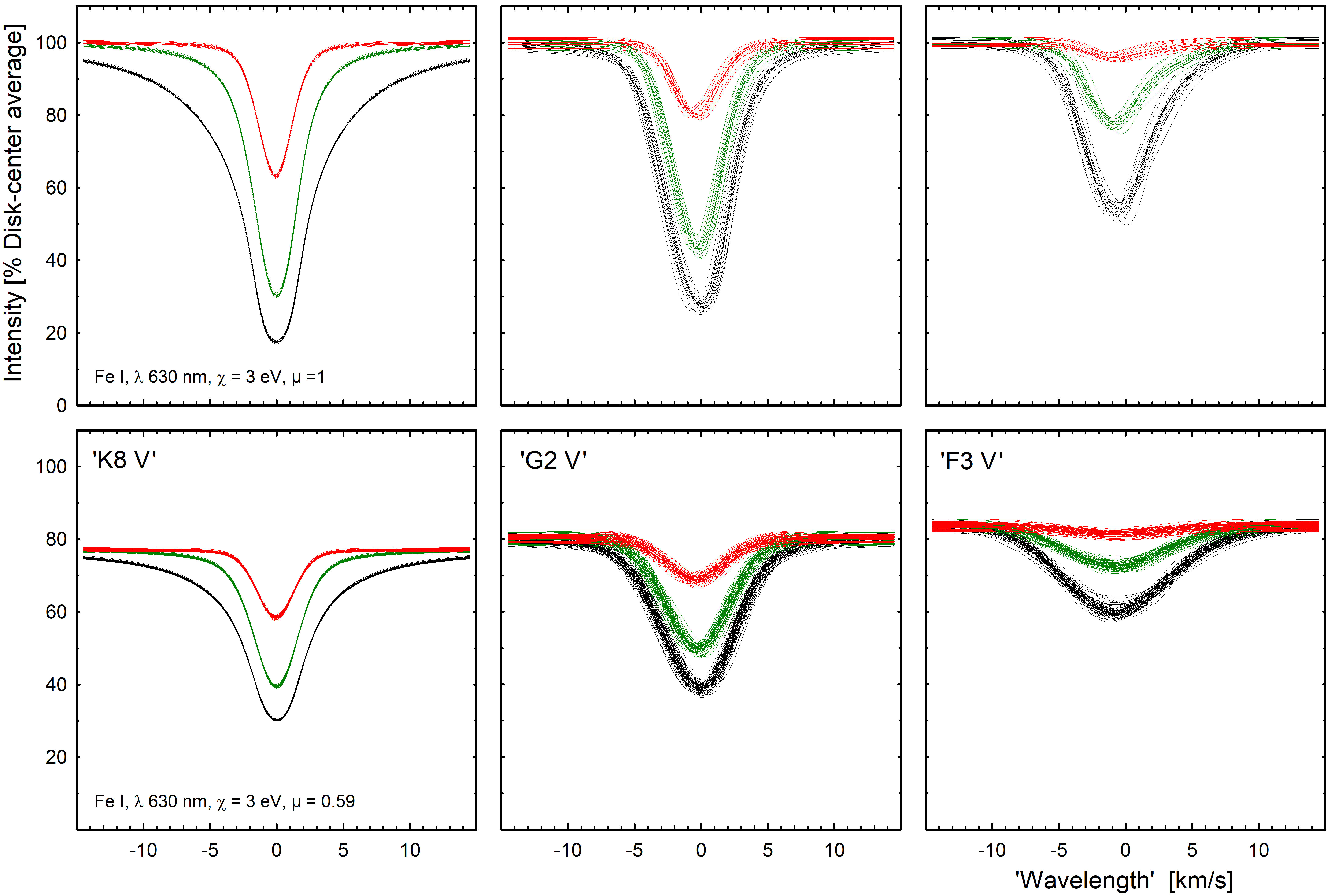}
     \caption{Spatially averaged but temporally resolved line profiles from (left to right) the `K8~V', `G2~V', and `F3~V' models at stellar disk center (top row) and at disk position $\mu$ = 0.59 (bottom).  Within each frame, spectral lines of three different strengths are shown: weak, medium and strong (red, green and black).  Profiles from some 20 instances in time during the modeling sequences are plotted for an idealized \ion{Fe}{i} line with $\chi$ = 3 eV at $\lambda$ 630 nm.  The number of profiles is greater for off-center data at $\mu$ = cos\,$\theta$ = 0.59, which at each instance in time include profiles at four different azimuth angles (quadrants on the star).  To discern the subtle variability in the `K8~V' simulation, this figure should be viewed highly magnified. } 
     \label{fig:temporal_profiles}
\end{figure*}

The discussion in this section, however, is limited to properties of single and idealized spectral lines.  Analogous to synthetic data presented in Papers I, II, and III, these are isolated \ion{Fe}{i} lines at $\lambda$ 630 nm, shown here for two different excitation potentials $\chi$ = 1 and 5 eV, and for three different line strengths.  These are the data entering Figs.\ \ref{fig:temporal_profiles}, \ref{fig:temporal_ranges}, and \ref{fig:temporal_correlations}.  Their parameters bracket those of typical lines in spectra of respective spectral types while not intended to model any specific real line.  With a single isolated line, its temporal response can be separated in a cleaner manner, without possible contamination from other line-formation effects (such as changing strengths of weak blending lines).  Of the wavering line properties, the jittering in radial velocity is the primary quantity that needs to be calibrated in the search for Earth-like exoplanets, and this is shown in particular for a solar model, for which observational tests should be possible using already existing observations \citep[e.g.,][]{dumusqueetal20}.

The character of temporal variability is illustrated in Fig.\ \ref{fig:temporal_profiles}.  Each superposed line profile for each of some 20 instances in time is formed as the average over thousands of spatial points across the simulation area (Table A.1 in Paper~IV).  These instances were selected to be sufficiently separated in time to ensure largely uncorrelated flow patterns between successive samples. These spatially averaged profiles do not (by far!) reveal the full spectral variability across stellar surfaces.  This full variability is substantial and spatially resolved profiles can be distorted into grossly asymmetric shapes and may occasionally even become double-bottomed when the trail of line formation has passed through differentially moving inhomogeneities \citep[e.g.,][]{asplund05, asplundetal00, bergemannetal19, chiavasssabrogi19, dravinsnordlund90}.  The categorization of these variegated features probably would be confusing, therefore we use sufficiently extended spatial averages, where the line profiles already start to closely resemble the final ones, and where ordinary fitting can be applied to obtain readily understandable quantities for line width, depth and wavelength shift.

Fig.\ \ref{fig:temporal_ranges} shows the jittering range of the wavelength shifts for differently strong lines in three models, at four different disk positions of $\mu$ = 1, 0.87, 0.59, and 0.21 as a function of the instantaneous irradiance from the simulation volume (normalized to the time-averaged flux at each respective stellar disk center).  For the solar model (`G2~V'), the fluctuation amplitudes tend to increase from disk center toward the limb, especially for the weak line, while in the hotter `F3~V' star, no such tendency is seen.  This is apparently connected with the transition from partially hidden granulation in the Sun to more naked granulation in F stars.  The full amplitude of solar granulation is encountered only very deep in the photosphere (even slightly below the visible surface), which is barely reached by even very weak lines in the optical.  With the convective patterns rapidly changing with height, lines formed under marginally different conditions may already show dissimilar signatures.  For F-type stars, with the full granulation contrast in optical view, these differential effects are much weaker, even though the convection as such is more vigorous.  In the cooler K-type stars, the full amplitude of granular convection largely remains hidden somewhat beneath the optically visible surface, and as a consequence, the convective wavelength shifts and their variations are much smaller.  In order to show their miniscule jittering, the scale for the `K8~V' star in Fig.\ \ref{fig:temporal_ranges} is expanded tenfold relative to that for the other stars.

\subsection{Proxies for jitter in radial velocity}

Searches for low-mass exoplanets would clearly benefit when some practically measurable quantities were identified that could be used to correct short-term excursions in apparent stellar radial velocity.  A considerable effort has been made in studying the Sun as a star and in searching for statistical relations between various parameters of solar and stellar variability, including work by \citet{ceglaetal19}, \citet{colliercameronetal19, colliercameronetal20}, \citet{debeursetal20}, \citet{dumusqueetal15, dumusqueetal20}, \citet{haywoodetal16}, \citet{lanzaetal16, lanzaetal19}, \citet{maldonadoetal19}, \citet{marchwinskietal15} and \citet{thompsonetal20}, and references therein.  These studies used various statistical approaches to search for possible correlations and empirical proxies for radial-velocity fluctuations.  Considerations of other spectral types include \citet{dumusqueetal11a} and \citet{meunieretal10, meunieretal17a, meunieretal17b}. 

Precision photometry from space reveals the presence of granulation as flickering in brightness, with both observed and modeled connections between its temporal power spectra and stellar spectral type.  In particular, flickering in brightness correlates with excursions in apparent radial velocity \citep{ceglaetal14, hojjatpanahetal20, oshaghetal17}.  However, securing space photometry simultaneously with spectroscopic radial-velocity surveys might be problematic although such photometry could select specific targets of particularly low flickering level, presumably with also low levels of radial-velocity jitter \citep{luhnetal20a, luhnetal20b, tayaretal19}.  

In this project, however, we did not start from full-disk correlations but instead from fine surface details on an ab initio level, with the aim to eventually extrapolate their properties to complete stellar disks.  Even if most starlight originates from thermally driven granulation, the present work is necessarily incomplete as it will require future modeling of much larger structures as well (supergranulation, etc.), and also an identification of the effects of granular magnetic regions and starspots. 

\begin{figure*}
\sidecaption
 \centering
 \includegraphics[width=12cm]{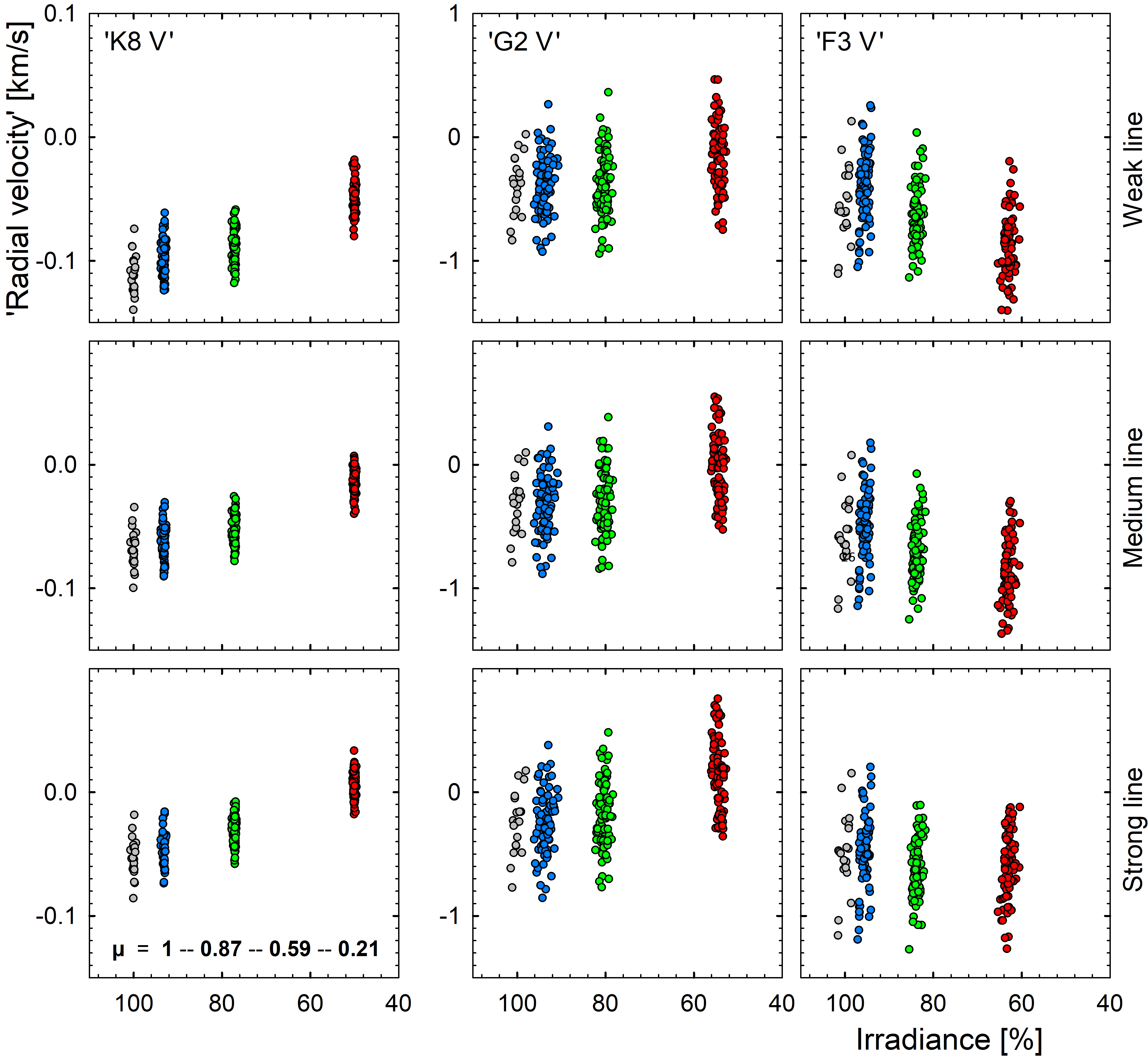}
     \caption{Range of jittering in wavelength shift for (left to right) the `K8~V', `G2~V', and `F3~V' models as a function of the surface irradiance at different times during each simulation. The velocity scale for the cool and quiet `K8~V' model is expanded tenfold relative to the others. This plot is (from top down) for idealized weak, medium, and strong \ion{Fe}{i} lines of $\chi$ = 3 eV at $\lambda$ 630 nm.  The four disk positions of $\mu$ = 1, 0.87, 0.59, and 0.21 are left to right inside each frame and marked in gray, blue, green and red. The number of points is four times greater for off-center positions than for $\mu$ = 1 because those are sampled at four different azimuth angles.  Irradiance is in units of the time-averaged disk-center continuum flux at $\lambda$ 630 nm in each model.  } 
     \label{fig:temporal_ranges}
\end{figure*}

Fig.\ \ref{fig:temporal_correlations} shows correlations for jittering in radial velocity and line depth, as function of fluctuations in irradiance.  These data from the `G2~V' model are for four different idealized synthetic \ion{Fe}{i} lines at $\lambda$ 630 nm: weak and strong, low- and high-excitation potentials ($\chi$ = 1 and 5 eV), at the two disk positions of $\mu$ = 1 and 0.87, which are representative for a large fraction of the full-disk stellar flux.  Each point in Fig.\ \ref{fig:temporal_correlations} originates from fitting the line profile averaged over the simulation area at some instance in time.  The four times greater number of points for $\mu$ = 0.87 reflects line profiles computed at four different azimuth angles, corresponding to observing four quadrants of the stellar disk.  Similar to Fig.\ \ref{fig:temporal_ranges}, irradiance is in units of the time-averaged disk-center continuum flux at $\lambda$ 630 nm (for an example of its temporal variability, see Fig.\ 2 of Paper I).   Because granular contrast changes with wavelength, the amplitude in irradiance must also depend on wavelength, potentially offering another diagnostic.  The exact amplitude will also depends on the simulation details (e.g., the oscillation modes that are treated) and its spatial extent (see Table A.1 in Paper IV).  

 Obvious correlations are present: increased irradiance enhances the convective blueshift and strengthens the spectral-line absorption (presumably caused by enhanced gradients in regions of line formation).  This is most pronounced for the strong line, and at stellar disk center.  The values in Figs.\ \ref{fig:temporal_ranges}-\ref{fig:temporal_correlations} are from Gaussian-type functions fit to line profiles, as described above.  More detailed correlation functions between fluctuating line shapes and wavelength shifts may be obtained by examining individual line profiles \citep{ludwigsteffen08}. 

An estimate of the trends indicates that excursions around $\pm$100 m\,s$^{-1}$ correspond to about $\pm$1\% in irradiance.  This applies to the tiny simulation patch of 5.6 Mm square (Table A.1 in Paper~IV).  The full solar disk is greater in effective area (recognizing the limb darkening) by a factor of $\sim$50,000, and when we assume random fluctuations and neglect center-to-limb variations, for the full disk these amplitudes would decrease by a factor $\gtrapprox$\,200 to about 0.5 m\,s$^{-1}$ per 50 ppm in irradiance.  These values are consistent with values deduced by \citet{meunieretal15} and \citet{meunierlagrange20} for the granulation-induced signal in apparent solar full-disk radial velocity although further contributions may come from larger-scale structures such as supergranulation \citep{meunierlagrange19}, which are not yet accessible to detailed 3D modeling.  A further modeling caution is that brightness variations are also driven by p-mode oscillations which are incompletely treated in the small simulation volumes.

The most important quality is that there do exist clear correlations between fluctuations in apparent radial velocity and other parameters, which can also be precisely measured from the ground.  In Fig.\ \ref{fig:temporal_correlations} this is shown for the line depth, normalized to the local spectral continuum.  (However, absolute line depths in units of the instantaneous stellar irradiance would require measurements from space; \citeauthor{ceglaetal19} \citeyear{ceglaetal19}.)  To firmly establish the quantitative correlations, simulations over more extended areas and times are required, and also an examination of which types of spectral lines (or combinations thereof) that are most compliant.  This may well differ for different stars.

From 3D simulations, not only line profiles but also broad-band photometric colors can be obtained.  Being less challenging than precision measurements of stellar irradiance, broad-band photometric colors can also be measured from the ground, although this requires a good understanding of atmospheric extinction.  Color changes reflect stellar temperature variations and carry information analogous to flickering in brightness, supplementing line depths as another proxy for excursions in radial velocity.  For CO\,$^5$BOLD models, synthetic colors were evaluated by \citet{bonifacioetal18} and \citet{kucinskasetal18}, and for the STAGGER grid by \citet{chiavassaetal18}.  Most probably, color indices spanning some particularly temperature-sensitive region (perhaps the G band around 415-440 nm, with its many molecular transitions) would provide the highest sensitivity.

\section{Future challenges} 

Just as on the Sun, most of the surfaces of solar-type stars is likely covered by thermally driven patterns of granular convection, producing the spectral features discussed above and in Paper~IV.  However, just as on the Sun, stars also  have some additional and more complex surface structures.  Spatially resolved spectroscopy should be able to help distinguish the effects from at least some of these.

\subsection{Stellar magnetic activity}

Spectra from stellar magnetic regions are expected to display specific characteristics.  On one hand, spectra retrieved from the surface regions that become temporarily hidden behind a large exoplanet during its transit will reveal the character of whatever stellar features then happened to lie behind the planet.  Assuming adequate S/N can be achieved, this would enable testing the modeling and simulations of various magnetic features on different stars.

On the other hand, however, stellar activity causes challenges for the calibration of radial-velocity jittering in searches for small exoplanets.  Even if satisfactory modeling of normal thermally driven granulation and fluctuations in its ensuing spectra is achieved, there remains a need to understand and segregate additional signals from time-variable magnetic activity.  Adequate modeling might permit us to identify spectral signatures that are specific to different classes of magnetic features and then enable their separation.  The quantitative level of the required effort strongly depends on the level of magnetic activity; for example, the Sun hosts numerous active regions during some periods but during others, it appears completely devoid of spots. 

For several stars, precision photometry shows instances when the transiting planet apparently crosses some dark spot \citep{brunoetal18, ioannidisschmitt16a, ioannidisschmitt16b, mancinietal17, mocniketal17, oshaghetal13, pontetal07, singetal11, valio13}, starspots can be searched for in various spectral signatures \citep{borradeschatelets15, morrisetal20b, reinersetal13}, and might perhaps be possible to glimpse with long-baseline optical interferometry \citep{ligietal15}. Furthermore, bright flares may appear in chromospheric emission lines \citep{czeslaetal15, klocovaetal17}.  The presence of emerging and decaying starspots complicates exoplanet discoveries \citep{barnesetal17, dumusqueetal11b, herreroetal16, isiketal18, lagrangeetal10, lanzaetal11, lisogorskyietal20, suarezmascarenoetal17, taloretal18}, and magnetic activity cycles also add to the complications \citep{lovisetal11}.  As we explained in Paper~I, one of the ambitions of the current project is to recover spatially resolved spectra also of starspots and stellar magnetic regions (including Zeeman signatures in magnetically sensitive spectral lines), which requires simultaneous precision photometry to identify the exact time of starspot transit. 

Here enters a certain limitation in how precisely the expected spectra can be modeled.  In the cases discussed so far of thermally driven convection, the basic stellar parameters of surface gravity and effective temperature suffice to formulate a hydrodynamic problem for which ab initio simulations can be carried out.  In the case of magnetically affected granulation, however, the unknown quantity of the total magnetic flux on (parts of) the stellar surface enters, which cannot be precisely measured (probably not even for the Sun).  In response to transiting active regions, the Sun (seen as a star) varies in irradiance \citep{toriumietal20} and in apparent radial velocity \citep{dumusqueetal20}.  It has been suggested that the Sun is less active than other G-type stars \citep{reinholdetal20}, although other arguments exist in the literature.  In any case, magnetic activity tends to decrease with age and with slower stellar rotation.

After introducing a certain magnetic flux in 3D atmospheric structures, its development can be followed into magnetic flux concentrations around intergranular lanes, the magnetic forces being balanced by gas pressure and dynamic effects.  In cooler stars with higher photospheric pressures, magnetic fields can be compressed to greater flux densities than is possible on the Sun.  This surface magnetoconvection, reproducing the appearance of distorted granulation in solar facular or plage regions, has been modeled, for example, by \citet{beecketal15a}, \citet{khomenkoetal18}, and \citet{salhabetal18}.  Spectral synthesis illustrates how the magnetic properties are reflected in the shapes of spectral lines, depending on their magnetic sensitivity \citep{holzreutersolanki12, holzreutersolanki15}.  For our project, the main relevance is for spectral-line properties at various center-to-limb distances for different classes of stars \citep{beecketal15b}. 
 
With sufficiently strong flux concentrations, surface convection is inhibited and starspots develop, whose flux densities and other properties seem determined by basic stellar parameters.  While the varying amount of magnetic flux for any one star may determine how many magnetic regions are present, it probably does not greatly affect properties of individual starspots, making them promising targets for spatially resolved spectroscopy because any single starspot should be a representative object.  Quite detailed 3D magnetohydrodynamic modeling of starspot umbrae and penumbrae is becoming possible for different stellar types \citep{panjaetal20}, and with spectral synthesis a logical next step, simulated data should soon become available, eventually to be compared with observed spectra of spatially resolved stellar features.  For precise determinations of magnetic field strengths, pairs of spectral lines with different Zeeman sensitivities may be used \citep{smithasolanki17}.

\begin{figure}
 \centering
 \includegraphics[width=\hsize]{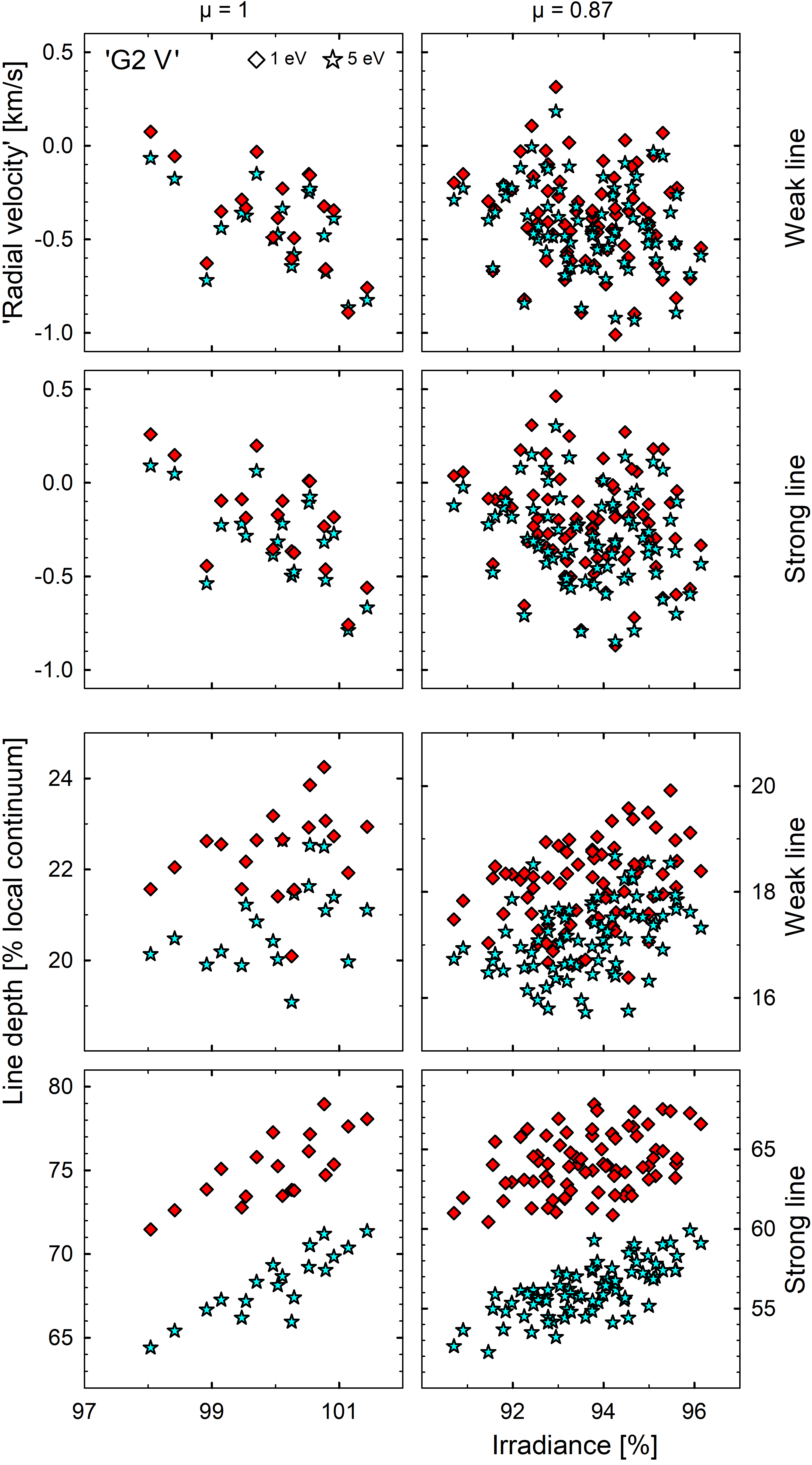}
     \caption{Correlations between flickering in irradiance and jittering of apparent radial velocity (top) and of absorption-line depth (bottom), for differently strong lines of both low and high excitation-potentials in a solar (`G2~V') model.  Measures for idealized \ion{Fe}{i} lines at $\lambda$ 630 nm are shown at stellar disk center ($\mu$ = 1) and at $\mu$ = 0.87; for $\chi$ = 1 eV (red diamonds) and 5 eV (blue stars).  The greater number of points for $\mu$ = 0.87 is provided by spectra at four different azimuth angles.  Irradiance is in units of the time-averaged disk-center continuum flux at $\lambda$ 630 nm. } 
     \label{fig:temporal_correlations}
\end{figure}

\subsection{Future observations}

Spatially resolved stellar spectroscopy requires high-fidelity spectra at high spectral resolution, with exceptionally low photometric noise, and a very stable wavelength calibration.  Data like this were not commonly available in the past, but can now be obtained at a great number of spectrometers that have recently been put in operation, or are being constructed for the primary task of radial-velocity searches for exoplanets.  The radial-velocity signal is obtained by cross correlating the observed cool-star spectrum with some template and the requirement for high spectral resolution follows from the need to preserve a high spectral-line contrast to produce a clear correlation signal.  To reach adequate S/N, an extended spectral coverage throughout the line-rich visual range is needed, combined with a wavelength stability striving for better than 1~m\,s$^{-1}$.  A few instruments operate in the near-infrared, although spectral fidelity may begin to become compromised by increased amounts of telluric absorptions.  Still, the infrared may be promising for identifying signatures from stellar magnetic regions because line splitting due to the Zeeman effect increases swiftly with wavelength.

These spectra can fulfill the requirements for spatially resolved spectroscopy, as was demonstrated in Papers~II and III.  Still, the spectral resolutions of almost all of these radial-velocity instruments are limited to $\sim$100,000.  As discussed above, to reveal more details of stellar photospheres, higher resolution is desirable, but there are very few suitable night-time instruments on large telescopes that also cover extended spectral ranges.  Here, ESPRESSO on the ESO VLT and PEPSI on the LBT stand out, with resolutions of about 200,000 and 300,000, which will hopefully be joined by HIRES on the ESO ELT in the future (with a planned resolution of 150,000 but with details still pending). 

For the illustrations in Fig.\ \ref{fig:noisy_reconstruction}, the area of transiting planets was assumed as 1.5\% of the stellar disk.  This is the observed value for the planet transiting the G0~V star HD\,209458 (Paper~II), while the planet that transits the smaller K1~V star HD\,189733A covers $\sim$2.5\% (Paper~III).  For smaller or larger planets, the required S/N scales inversely with planet area.  Even if many spectral lines and many recordings can (and must) be averaged, perhaps from multiple transits, this stacking decreases the random noise only by the square root of the number of exposures, a slowly growing function.  In practice, this limits realistic targets to host starts with transiting planets covering at least 0.5\%, for instance, of their disks.  

Although numerous stars are known with transiting planets of this size (Fig.\ 1 in Paper~II), not many are also bright enough to permit truly low-noise spectra to be recorded.  However, ongoing searches for bright host stars with transiting exoplanets continue to find new targets spanning rather broad temperature ranges \citep[][etc.]{bakosetal20, grievesetal21, kaneetal20, lundetal17, mancinietal13, martinezetal20, pepperetal13, pepperetal20, schrijver20, wangetal19}.  In addition to apparent stellar brightness, the suitability of a target depends on the line richness of the spectrum, and how wide a spectral range can be covered.  Multiple spectral lines will have to be averaged but the extent of this averaging (as in Papers~II and III) is constrained by the number and types of spectral lines that are present.  Certainly, more lines than merely from Fe in the visual can be used, but if reasonably unblended lines are required from atomic species without complications of pronounced isotopic or hyperfine structure, their total number may rather be in the hundreds, not thousands.  For the solar spectrum, a survey of lines from these common species (\ion{Fe}{i}, \ion{Fe}{ii}, \ion{Ti}{i}, \ion{Ti}{ii}, etc.) is presented in \citet{dravins08}.  A final constraint is set by the number of detectable photons \citep{bouchyetal01, reinerszechmeister20}, which means that these studies will greatly benefit by extremely large telescopes.

\subsection{Conclusions and outlook}

This survey of simulations of realistically complex spectra for stars both cooler and hotter than the Sun has identified specific 3D signatures in various categories of spectral lines that reflect the dynamic fine structure in their photospheres.  This is where the spectral lines originate, whose wavelength shifts are used to infer stellar radial motion and its possible modulation by exoplanets.  Current simulations treat the thermally driven surface granulation, which just as on the Sun is expected to cover most of the surface of solar-type stars and contribute the largest fraction by far of the spectral flux.  The validity of these models can be tested with spatially resolved stellar spectroscopy, where stellar line profiles are retrieved from behind transiting exoplanets.  While ordinary granulation produces the main patterns of convective wavelength shifts, additional contributions are expected from large-scale structures such as supergranulation or meridional flows, and from regions of magnetically affected granulation as well as starspots.  Although large-volume stellar simulations are challenging, they are expected eventually to be able to model supergranulation as well, while spatially resolved spectroscopy of starspots will be enabled when some exoplanet happens to transit them. 

One conclusion is that when the spectral resolution $\lambda$/$\Delta\lambda$ reaches $\sim$100,000, the instrumental degradation of the overall line parameters of width, depth and wavelength position is substantially smaller than the expected physical changes across stellar disks (Fig.\ \ref{fig:resolution_parameters}).  This enables the testing of atmospheric 3D models by spatially resolved spectroscopy already from the fitting of somewhat noisy data, as also demonstrated in Papers~II and III.  However, distinguishing the altering bisector shapes between different classes of spectral lines in any one star appears to require extremely high S/Ns (Fig.\ \ref{fig:noisy_reconstruction}), likely requiring data from future extremely large telescopes.

The temporal variability of spectral line parameters can be predicted from such ab initio simulations.  Because different parameters vary together, proxies can be identified for the jittering of apparent radial velocity.  When these quantities can be practically measured from the ground, they should provide a means to adjust measured velocities to true stellar radial motion.  In particular, the jittering in apparent radial velocity correlates with fluctuations in line depth and of photometric color (Figs.\ \ref{fig:temporal_ranges}-\ref{fig:temporal_correlations}).  With a sufficient physical understanding of the detailed line formation processes in stellar photospheres, it should eventually become possible to also identify truly Earth-like planets.

\begin{acknowledgements}
{The work by DD is supported by grants from The Royal Physiographic Society of Lund.  HGL gratefully acknowledges financial support by the Deutsche Forschungsgemeinschaft (DFG, German Research Foundation) -- Project ID~138713538--SFB~881 (`The Milky Way System', subproject A04).  Use was made of NASA’s ADS Bibliographic Services and the arXiv$^{\circledR}$ distribution service.  We thank the referee, Andrea Chiavassa, for his detailed and insightful comments.  }

\end{acknowledgements}

%\normalsize 

\end{document}